\documentstyle[11pt]{article}

\begin{document}

\thispagestyle{empty}

\begin{center}
{\Large \bf Noncommutative Integrability \\and Recursion Operators}
\footnote[1]
{Supported in part by the italian Ministero
dell' Universit\`a e della Ricerca Scientifica e Tecnologica.\newline
Corresponding author: G. Vilasi, Dipartimento di Scienze Fisiche ''E. R.
Caianiello'', via S. Allende, 84081-Baronissi (Salerno), Italy.\newline
Phone: 039-089-965317; Fax: 039-089-965275; e-mail: vilasi@sa.infn.it}

\end{center}

\vspace{1.5cm}

\begin{center}
{\Large Giovanni Sparano $^{{\#,\S }}$, \ Gaetano Vilasi $^{{\#,\S \S }}$ }

\end{center}

\bigskip

\begin{center}
{$^{{\#}}${\it Erwin Schr\oe dinger International Institute\ for
Mathematical Physics, }\\
{\it Boltzmanngasse 4, A-1090 Wien, Austria.}\\
$^{{\#}}${\it Istituto Nazionale di Fisica Nucleare,
Gruppo Collegato di Salerno, Italy.}\\
$^{{\S }}${\it Dipartimento di Matematica e
Informatica, Salerno University, Italy. }\\
$^{{\S \S }}${\it Dipartimento di Scienze Fisiche
''E.R.Caianiello'',Salerno University, Italy}\\}
\end{center}
\bigskip

\vspace{2.5cm}

\begin{abstract}
Geometric structures underlying commutative and non commutative integrable
dynamics are analyzed.
They lead to a new characterization of noncommutative integrability in
terms of spectral properties
and of Nijenhuis torsion of an invariant (1,1) tensor field. The
construction of compatible symplectic
structures is also discussed.
\par
{\it Subj. Class.}: Dynamical Systems.
\end{abstract}

\vspace{8.0cm}

\vfill\eject

\section{Introduction}

In the last few decades there has been a renewed interest in completely
integrable Hamiltonian systems, whose concept goes back to the last century
\cite{Lio855} and which, loosely speaking, are dynamical systems admitting a
Hamiltonian description and possessing sufficiently many constants of motion,
so that they can be integrated by quadratures. Some qualitative features of
these systems remain true in some special classes of infinite--dimensional
Hami\-l\-tonian systems ex\-pressed by nonlinear evolution equa\-tions as, for
instance, Korteweg-de Vries and sine-Gordon \cite{Vi80}.

A relevant progress in the study of these systems with an infinite-dimensional
phase manifold $\cal{M}$, was the introduction of the {\it Lax
Representation }\cite{La68} which played an important role in formulating the
{\it  Inverse Scattering Method}, universally recognized as one of the most
remarkable result of theoretical physics in last decades, and of the
{\it  AKNS scheme} \cite{AKNS74}. This method allows the integration of non
linear dynamics, both with a finitely or infinitely many degrees of freedom,
for which a Lax representation can be given \cite{GGKM67}, this being both of
physical and mathematical relevance \cite{FT87}.

Most of \ the evolution equations admitting a Lax Representation are generally
Hamiltonian dynamics on infinite dimensional {\it  weakly-symplectic}
manifolds, so that the natural arena, for the analysis of their integrability,
is represented by the phase space with its natural symplectic structure. In
terms of this structure, the scattering data associated to the Lax operator
have a natural interpretation as {\it  action-angle} type variables
\cite{ZF71}.

A further progress, in the analysis of the integrability, was the important
remark that many of previous systems are Hamiltonian dynamics with respect to
two {\it  compatible} symplectic structures \cite{Ma78,Ma80,GD80,Vi80}, this
leading to a geometrical interpretation of the so called {\it  recursion
operator }\cite{La68}{\it  .}

This fact suggested that the integrability of non linear field theories could
be naturally explained in terms of mixed tensor fields, whose relation
\cite{La68} with Lax operators is still \cite{DMSV82} unclear.

As a matter of fact, a description of integrability
\cite{Ma80,DMSV82,FY83,DMSV84,DSV85,KS86,LMV94}, which does not depend in a
crucial way on
dimensionality, and so works both for systems with finitely many degrees of
freedom and for field theory can be given in terms of invariant mixed tensor
field, having bidimensional eigenspaces and vanishing Nijenhuis torsion
\cite{Ni51}.

The analysis of the integrability realized with the help of a such tensor
field leads to the formulation of an integrability criterion
\cite{DMSV82,DMSV84,DSV85,LMV94} which, for finite-dimensional systems, is
{\it  essentially} equivalent to the classic Liouville theorem.

To be more specific, the mentioned {\it  essential equivalence }means that
the equivalence holds for {\it  non resonant} Hamiltonian systems,
{\it  i.e.,} for completely integrable dynamics whose Hamiltonian expressed
in action-angle coordinates has a non vanishing Hessian.

One reason for a completely integrable Hamiltonian system to be resonant may
be that the number of first integrals, defined on the entire phase space, is
larger than one half of the phase space dimension (of course, in this case not
all the integrals are in involution and one will have to deal with
noncommuting sets of first integrals). This happens for the Kepler dynamics
which, however, is bihamiltonian and has a recursion operator with the right
properties \cite{MV92}.

More in general, the analysis of symmetries \cite{KLV86} shows that generally
one is faced with a non Abelian algebra corresponding, for Hamiltonian
systems, to a non Abelian algebra of first integrals.

The integrability of such systems, with finitely many degrees of freedom, has
been analyzed in several papers \cite{MF78}. There exist field dynamics,
related to vector and matrix nonlinear Schroedinger equation \cite{Ku85,Ge94},
possessing a noncommutative set of first integrals so that it would be useful
to have a noncommutative integrability criterion formulated in terms of a
recursion operator. In this paper such a criterion is presented. More
specifically, in section 2 integrability criteria for the commutative case are
recalled. Section 3 is devoted to noncommutative integrability. After a review
of known results a new noncommutative integrability criterion is presented.
The Kepler dynamics is discussed as an example.

\section{Commutative integrability criteria}

The best known criterion of integrability goes back to the celebrated
Liouville theorem \cite{Lio855} and can be reported
\cite{Ar76,VK77,AM78,AKN88}
as follows:

\vskip.5cm

\noindent{\bf Theorem}
{\it If on a $2n$ dimensional symplectic ma\-nifold ${\cal M}$ are de\-fined a
Ham\-iltonian dynamics and $n$\ functionally independent first integrals
$f_{1},\dots,f_{n}$\ in involution
\[
\left\{  f_{i},f_{j}\right\}  =0
\qquad
\forall i,j=1,\dots,n\,\,\,,
\]
whose associated Hamiltonian fields $X_{i}$ are complete, then the level
manifolds
\[
{\cal M}_{f\left(\pi\right)}=\left\{p\in{\cal M}:f_{i}\left(p\right)  =
\pi_{i}\qquad i=1,\dots,n\right\}
\]
are invariant with respect to the dynamics and each of their connected
components is diffeomorphic either to $T^{m}\times\Re^{n-m}$or, if compact, to
a torus $T^{n}$. Moreover, for every point $p\in{\cal M}$ near which $m$ is
constant, there exists a neighborhood ${\cal U}$ invariant under the
composed flow of the vector fields $X_{i}$, and canonical coordinates $\left(
P_{1},\cdots,P_{n},Q^{1},\cdots,Q^{n}\right)  $, where $Q^{1},\cdots,Q^{m}$
are angles, such that the equation of the motion take the form:
\[
\dot{P}_{i}=0,\quad\,\dot{Q}^{i}=\nu^{i}\left(  P\right)  ,\,\,\,\,1\leq
\,\,i\leq n.
\]
}
A more general setting for the commutative integrability is the
following\ \cite{LMV94,Vi99}.

Let ${\cal M}$ be a smooth $2n$-dimensional manifold. Let us suppose we can
find $n$ vector fields $X_{1},\dots,X_{n}\in{\cal X}({\cal M})$ and $n$
functions $f_{1},\dots,f_{n}\in{\cal F}({\cal M})$ with the following
properties
\begin{eqnarray}
& [X_{i},X_{j}]=0~,\label{eq:cis1}\\
& L_{X_{i}}f^{j}=0~.~~~i,j\in\{1,\dots,n\}~.\label{eq:cis2}%
\end{eqnarray}
It can be shown that, if on an open dense submanifold of ${\cal M}$
\begin{eqnarray}
& X_{1}\wedge\cdots\wedge X_{n}\not =0~,\label{eq:cis4}\\
& df^{1}\wedge\cdots\wedge df^{n}\not =0~,\label{eq:cis5}%
\end{eqnarray}
any dynamical system $\Delta$ on ${\cal M}$ which is of the form
\begin{equation}
\Delta=\sum_{i=1}^{n}\nu^{i}X_{i}~,~~~\nu^{i}=\nu^{i}(f^{1},\dots
f^{n})~,\label{eq:cis3}%
\end{equation}
is completely integrable on the submanifold on which Eqs. (\ref{eq:cis4}) and
(\ref{eq:cis5}) are satisfied.

If the fields $X_{i}$ are complete, by using the $n$-functions $f^{1}%
,\dots,f^{n}$ , a family of symplectic structures can defined with respect to
which the dynamics is Hamiltonian.

In the Liouville theorem \cite{Lio855} only the commuting first integrals and
the symplectic structure are given. Of course, the commuting vector fields are
constructed from them. An alternative integrability theorem, suggested by the
analysis of integrable models in field theory, can be formulated
\cite{DMSV82,DMSV84,DSV85} using invariant tensor fields and it reads :

\vskip.5cm

\noindent{\bf Theorem (DMSV)}
{\it Let $\Delta$ be a dynamical vector field on a differential manifold
${\cal M}$ which admits a $\left(  1,1\right)  $ mixed tensor field $T$ which

\begin{itemize}
\item {\it  is invariant}
\[
L_{\Delta}T=0
\]

\item {\it  has a vanishing Nijenhuis torsion}
\[
{\cal N}_{T}=0
\]

\item {\it  is diagonalizable with doubly degenerate eigenvalues }%
$\lambda_{j}$ {\it  whose differentials }$d\lambda_{j}$ {\it  \ are
independent at each point }
\end{itemize}

Then, the vector field $\Delta$\ is separable, completely integrable and
Hamiltonian{\it  .}
}
\vskip.5cm

\noindent{\bf Remark}
{\it
We observe that the Hamiltonian character of the dynamics $\Delta$ is not
assumed {\it  a priori} but it follows from the properties of the tensor
field $T$, so that all dynamics, satisfying the given hypotheses, result to be
Liouville integrable. Integrability of dissipative dynamics can be put in the
same setting by assuming \cite{DMSV83} different spectral hypothesis for the
tensor field $T$.
}
The last formulation has the advantage of being more appropriate to deal with
dynamics with infinitely many degrees of freedom (completely integrable field
theories). We also observe that the Lax Representation, the powerful
integration tool for such systems, may not be useful in more than one space
dimension since the inverse problem in Quantum Mechanics has been solved only
for $1$-dimensional systems.

\subsection{From the Liouville integrability to invariant mixed tensor
fields.\label{inv}}

Let us now study the problem of constructing invariant mixed tensor fields,
with the appropriate properties (also called a {\it  recursion tensor
field}), for a given Liouville's integrable Hamiltonian dynamics $\Delta$. If
$H$ is the Hamiltonian function and $\left\{  \cdot,\cdot\right\}  $ denotes
the Poisson bracket, we have{\it  \ }
\[
\Delta f=\{H,f\}\,\,\,.
\]

Let us introduce in some neighborhood of a Liouville's torus $T^{n}$
action-angle coordinates $\left(  J_{1},...,J_{n},\varphi^{1},...,\varphi
^{n}\right)  $, in which we have:
\[%
\begin{array}
[c]{l}%
\omega=\sum_{h}dJ_{h}\wedge d\varphi^{h}\\
\displaystyle
\Delta=\frac{\partial H}{\partial J_{h}}\frac{\partial}{\partial\varphi^{h}%
}\,\,\,.
\end{array}
\]

Let us distinguish two cases:

\begin{itemize}
\item  The Hamiltonian $H$ is a separable one
\[
H=\sum\nolimits_{k}H_{k}(J_{k})\,\,\,\,.
\]

In this case a class of recursion tensor fields can be easily defined
\[
T=\sum_{h}\lambda_{h}(J_{h})(dJ_{h}\otimes\frac{\partial}{\partial J_{h}%
}+d\varphi^{h}\otimes\frac{\partial}{\partial\varphi^{h}})
\]
with the $\lambda$'s arbitrary and functionally independent. Indeed, the
tensor field $T$ is invariant and has vanishing Nijenhuis torsion and doubly
degenerate eigenvalues.

\item  The Hamiltonian has a non vanishing Hessian:
\[
\det\left(  \frac{\partial^{2}H}{\partial J_{h}\partial J_{k}}\right)  \neq0
\qquad
.
\]

In this case, in the chosen neighborhood, setting
\[
\nu^{h}(J)=\frac{\partial H}{\partial J_{h}},
\]
new coordinates $\left(  \nu/\varphi\right)  $ can be introduced, so that the
dynamics can be described, with respect to the new symplectic structure
\[
\omega_{1}=\sum_{h}d\nu^{h}\wedge d\varphi^{h}=\sum_{hk}\frac{\partial^{2}%
H}{\partial J_{h}\partial J_{k}}dJ_{k}\wedge d\varphi^{h},
\]
by a separable Hamiltonian function:
\[
H_{1}={\frac{1}{2}}\sum_{h}(\nu^{h})^{2}\,\,\,.
\]

As before, a class of recursion tensor fields is then given by
\[
T=\sum\nolimits_{h}\lambda_{h}(\nu^{h})(d\nu^{h}\otimes\frac{\partial
}{\partial\nu^{h}}+d\varphi^{h}\otimes\frac{\partial}{\partial\varphi^{h}%
})\,\,\,\,.
\]
\end{itemize}

By means of this construction it is possible to find the second symplectic
structure for a completely integrable Hamiltonian system.

It is still an open problem if this is true also in the remaining cases. In
this direction one may find useful hints in \cite{Bo96,FR98}.

Next section is concerned with noncommutative integrable dynamics and, in
particular, with their characterization in terms of an invariant, mixed tensor
field.

\section{Noncommutative integrability criteria.}

As it has been observed in the Introduction, if the number of independent
first integrals is larger than half the dimension of the symplectic manifold,
they cannot be in involution anymore and one will have to deal with
noncommuting sets of first integrals. For a finite number of degrees of
freedom a noncommutative generalization of Liouville theorem is the following
\cite{MF78,AR88}:

\vskip.5cm

\noindent{\bf Theorem (MF1)}
{\it A Hamiltonian vec\-tor field on a sym\-plectic
man\-i\-fold $\left(
{\cal M},\omega\right) $ having a noncommutative Lie algebra ${\cal A}
$ of first integrals satisfying the condition
\[
dim {\cal A}+rank {\cal A} = dim {\cal M},
\]
where $n$-$rank{\cal A}$ is the maximum of the rank of the
matrix $m_{ab}=\left\{  f_{a},f_{b}\right\}  ${\it  \footnote{For semisimple
Lie algebras, this definition coincides with the usual one.}}, is completely
integrable, i.e. the joint level surfaces of the first integrals are
invariant, and in a neighborhood of each invariant surface one can define
canonical coordinates $\left(  \lambda/\chi/p/q\right)  $, the $\chi$'s being
the coordinates on the invariant surfaces, such that Hamilton's equations take
the form
\[
\dot{\lambda}_{i}=0,\;\dot{\chi}^{i}=\nu_{i},\;\dot{p}_{\alpha}=0,\,\dot
{q}^{\alpha}=0,\,\,\,\,1\leq\,\,i\leq r,\quad\,\,r+1\leq\alpha\leq n,
\]
with $r=rank{\cal A}$. If these invariant surfaces are compact and
connected one can prove, as in the commutative case, that they are tori, and
the $\chi$'s can be chosen to be angle variables. The canonical coordinates
are called, in this case, ''generalized action- angle variables{\it  ''}.
}
The Liouville theorem can be recovered \cite{AKN88} as stated by:

\vskip.5cm

\noindent{\bf Theorem (MF2)}
{\it If ${\cal M}$ is compact, then, under the hypotheses of the previous
theorem, one can find $n=\frac{1}{2}\dim{\cal M}$ first integrals which are
in involution.
}
Even in this case, however, the noncommutative theorem, showing the full
symmetry of the system, remains of interest.

A full account of the relevant geometrical structures underlying the
noncommutative integrability, can be found in \cite{Mi95,FR98,Bo98}. Here we
just give a short review of them.

A symplectic form on ${\cal M}$ at a point $p$ defines a skewsymmetric
bilinear non degenerate form on $T_{p}{\cal M}$. If $W$ is a $r$%
-dimensional subspace of $T_{p}{\cal M}$, the symplectic orthogonal
subspace $W^{\bot}\equiv\left\{  X\in T_{p}{\cal M}:\omega\left(
X,Y\right) = 0 \qquad \forall Y\in W\right\}$ has dimension $2n-r$ and in
general $W\cap W^{\bot}$ $\neq$ {0}.

Two cases are of particular interest: $r\leq n$ and $r\geq n$. If $r\leq n$
and $W$ $\subseteq$ $W^{\bot}$, $W$ is said to be {\it  isotropic}; if
$r\geq n$ and $W\supseteq$ $W^{\bot}$, $W$ is called {\it  coisotropic}. If
$W$ is isotropic and coisotropic ($r=n$) then it is called {\it  Lagrangian}%
. A submanifold is called {\it  isotropic, coisotropic }%
or{\it  \ Lagrangian} if its tangent spaces are respectively isotropic
coisotropic or Lagrangian.

In the commutative case the level surfaces of the first integrals $f_{i}$
define an invariant Lagrangian foliation ${\cal F}_{1}$ of ${\cal M}$.
The Hamiltonian vector fields $X_{i}$ associated to the functions $f_{i}$ are
then a basis of commuting tangent vector fields for the leaves and can be used
to define local coordinates $\chi^{i}$ on the leaves. These fields also
commute with the Hamiltonian vector field $\Delta$ which, consequently, can be
expressed as $\Delta=\nu^{i}\left(  f\right)  X_{i}$. In a neighborhood of a
point $p\in{\cal M}$, the set $\left(  \chi/f\right)  $ define canonical
coordinates and Hamilton's equations of motion take the simple following
form:
\[
\dot{\chi}^{i}=\nu^{i},\,\,\,\dot{f}_{i}=0\,.
\]

In the noncommutative case the first integrals $f_{a}$, $\ \ 1\leq a\leq2n-r
$, still define an invariant foliation, but the leaves now have dimension
$r\leq n$ and the Hamiltonian vector fields $X_{a}$, associated with the first
integrals $f_{a}$, are not all tangent to the leaves. However, the condition
$dim {\cal A}+ rank {\cal A} = dim {\cal M}$ ensures, for each
leaf $l$, the existence of a subalgebra ${\cal A}_{l}$ which commutes with
${\cal A}$ on $l$. The Hamiltonian vector fields $\overline{X_{i}}$,
associated to a basis of ${\cal A}_{l}$, will give themselves a basis of
tangent vector fields for $l$ and will have the property $\left.
\omega\left(  \overline{X_{i}},X_{a}\right)  \right|  _{l}=0$, so that each
leaf will be \footnote{In particular $\left.  \omega\left(  X_{i}%
,X_{j}\right)  \right|  _{l}=0$.} isotropic. To obtain a set of canonical
coordinate, in a neighborhood of a point of \ $l$ and eventually of the whole
of $l$, one needs to exploit further properties of this isotropic foliation.
At each point $p$ of $l$ consider the subspace $T_{p}l$ $\subseteq$
$T_{p}{\cal M}$ and the resulting distribution of symplectically orthogonal
subspaces $\left(  T_{p}l\right)  ^{\bot}$.\ \ Since $\left.  \omega\left(
\overline{X_{i}},X_{a}\right)  \right|  _{l}=0$, this distribution is
generated, for all leaves, by the vector fields $X_{a}$, and, furthermore,
since $X_{a}$ satisfy the hypotheses of the Frobenius theorem, we obtain a
second coisotropic foliation ${\cal F}_2$ whose leaves are themselves
foliated by those of the first foliation ${\cal F}_1$. The regularity of
this foliation follows from the independence of the functions $f_{a}$. One can
now prove the existence of canonical coordinates $\left(  \lambda_{i},\chi
^{i},p_{\alpha},q^{\alpha}\right)  $, such that the symplectic structure and
the dynamical vector field take the following form
\[
\omega=d\lambda_{i}\wedge d\chi^{i}+dp_{\alpha}\wedge dq^{\alpha}%
,\,\,\,\Delta=\nu^{i}\left(  \lambda\right)  X_{i}%
\]
so that the equations of motion become
\[
\dot{\lambda}_{i}=0,\;\dot{\chi}^{i}=\nu^{i},\;\dot{p}_{\alpha}=0,\,\dot
{q}^{\alpha}=0.
\]

The functions $\ \lambda_{i}$ describe locally ${\cal F}_{2}$, and their
associated Hamiltonian vector fields $X_{i}$ define coordinates $\chi^{i}$ on
${\cal F}_{1}$. The fields $X_{i}$ are independent and, since
$\omega\left(  X_{i},X_{a}\right)  =d\lambda_{i}\left(  X_{a}\right)  =0$,
they are tangent to the leaves of ${\cal F}_{1}$, and thus commute among
themselves and with $\Delta$. To understand better this canonical coordinates,
one can actually observe that the momentum map $J:{\cal M}\rightarrow
{\cal A}^{\ast}$ defined by $J:x\rightarrow\xi_{x}\in$ ${\cal A}^{\ast}$
where $\xi_{x}(f)\equiv f(x)$, $f\in$ ${\cal A}$, defines a fibration of a
neighborhood ${\cal U}$ of a leaf of ${\cal F}_{2}$ with fiber
$l_{x}=J^{-1}\left(  \xi_{x}\right)  $, namely a leaf of ${\cal F}_{1}$.
The neighborhood ${\cal U}$ can then be represented as $l_{x}\times
S\times{\cal O}$, where ${\cal O}$ is a region in the coadjoint orbit through
$\xi_{x}$ of the Lie group corresponding to ${\cal A}$ and $S$ is a linear
manifold transverse to ${\cal O}$. The symplectic structure $\omega$
restricted to ${\cal O}$ coincides with the Lie-Kirillov-Kostant-Souriau
symplectic form; $\left(  p_{\alpha},q^{\alpha}\right)  $ are canonical
coordinates on ${\cal O}$ and $\lambda_{i}$ coordinates on $S$. It has been
actually proved \cite{FR98} that all what is needed for the existence of such
local canonical coordinates is the double foliation, namely that ${\cal M}$
has an isotropic foliation such that the distribution of subspaces,
symplectically orthogonal to the tangent spaces to its leaves, is integrable.

\subsection{ Noncommutative in\-tegrability and in\-variant tensor field.}

Let us give now the following new characterization of noncommutative
integrability.

\vskip.5cm

\noindent{\bf Theorem}
{\it Let $\Delta$ be a dynamical vector field on a $2n$-dimensional
manifold ${\cal M}$ which admits a $\left(  1,1\right)  $ mixed tensor
field $T$ which

\begin{itemize}
\item {\it  is invariant}
\[
L_{\Delta}T=0
\]

\item {\it  is diagonalizable with} only simple and {\it  doubly
degenerate \ eigenvalues} {\it  whose differentials are independent at each
point }$p\in{\cal M}$.

\item {\it  has the property}
\[
{\cal N}_T\left(  \alpha,X,Y\right)  =0
\]
$\forall X:X\left(  p\right)  \in S\left(  p\right)  $, $\forall
Y\in{\cal D}\left(  {\cal M}\right)  $ {\it  and for all }%
$1$-{\it  forms }$\alpha$, $S\left(  p\right)  $ {\it  denoting the sum of
eigenspaces associated to the doubly eigenvalues }of $T\left(  p\right)  $.
\end{itemize}

Then, the vector field $\Delta$\ is separable, completely integrable and
Hamiltonian{\it  .}
}
Let $\lambda_{1},\lambda_{2},..,\lambda_{r}$ be the doubly degenerate
eigenvalues and $\mu_{2r+1},...,\mu_{2n}$ be the simple ones. Under the
hypotheses, the tensor field $T$ can be written in the form
\begin{equation}
T=\sum_{i=1}^{r}\lambda_{i}\left(  e_{i}\otimes\vartheta^{i}+e_{i+r}%
\otimes\vartheta^{i+r}\right)  +\sum_{\alpha=2r+1}^{2n}\mu_{\alpha}e_{\alpha
}\otimes\vartheta^{\alpha},
\end{equation}
where the $e$'s form a basis of eigenvectors of $T$ and the $\vartheta
^{\prime}s$ are the elements of the dual basis. Thus,
\begin{equation}%
\begin{array}
[c]{llll}%
Te_{i}=\lambda_{i}e_{i}, & Te_{i+r}=\lambda_{i}e_{i+r}, & Te_{\alpha}%
=\mu_{\alpha}e_{\alpha}, & i\leq r,\quad\alpha\geq2r+1\\
T\vartheta^{i}=\lambda_{i}\vartheta^{i}, & T\vartheta^{i+r}=\lambda
_{i}\vartheta^{i+r}, & T\vartheta^{\alpha}=\mu_{\alpha}\vartheta^{\alpha}, &
i\leq r,\quad\alpha\geq2r+1.
\end{array}
\end{equation}

The Nijenhuis torsion \cite{Ni51} of $T$, defined by
\begin{equation}
{\cal N}_{T}\left(  \alpha,X,Y\right)  {\cal =}\left\langle
\alpha,{\cal H}_{T}\left(  X,Y\right)  \right\rangle
\end{equation}
with
\begin{equation}
{\cal H}_{T}\left(  X,Y\right)  =\left[  TX,TY\right]  +T^{2}\left[
X,Y\right]  -T\left[  TX,Y\right]  -T\left[  X,TY\right]  ,\label{bII,4,9}%
\end{equation}
once evaluated on the basis vector fields $\left\{  e_{1},\cdots
,e_{2n}\right\} $, gives
\begin{eqnarray*}
{\cal H}_{T}\left(  e_{i},e_{j}\right)   & =\left(  T-\lambda_{i}\right)
\left(  T-\lambda_{j}\right)  \left[  e_{i},e_{j}\right]  +\left(  \lambda
_{i}-\lambda_{j}\right)  \left[  \left(  L_{e_{i}}\lambda_{j}\right)
e_{j}+\left(  L_{e_{j}}\lambda_{i}\right)  e_{i}\right]  \,\,\,\\
{\cal H}_{T}\left(  e_{i},e_{\alpha}\right)   & =\left(  T-\lambda
_{i}\right)  \left(  T-\mu_{\alpha}\right)  \left[  e_{i},e_{j}\right]
+\left(  \lambda_{i}-\mu_{\alpha}\right)  \left[  \left(  L_{e_{i}}\mu
_{\alpha}\right)  e_{\alpha}+\left(  L_{e_{\alpha}}\lambda_{i}\right)
e_{i}\right]  \,\,
\end{eqnarray*}
where $i,j\leq2r$ and $\alpha\geq2r+1$, so that the conditions on the torsion
imply the following relations:
\begin{equation}%
\begin{array}
[c]{ll}%
\left(  T-\lambda_{i}\right)  \left(  T-\lambda_{j}\right)  \left[
e_{i},e_{j}\right]  =0, & \left(  \lambda_{i}-\lambda_{j}\right)  e_{i}\left(
\lambda_{j}\right)  =0,\\
\left(  T-\lambda_{i}\right)  \left(  T-\mu_{\alpha}\right)  \left[
e_{i},e_{\alpha}\right]  =0, & e_{i}\left(  \mu_{\alpha}\right)  =e_{\alpha
}\left(  \lambda_{i}\right)  =0
\end{array}
\label{bII,4,10}%
\end{equation}

It follows that for any three vector fields $e_{i}$, $e_{j}$,$e_{\alpha}$
\[%
\begin{array}
[c]{cc}%
\left[  e_{i},e_{j}\right]  =ae_{i}+be_{j}+ce_{i+r}+de_{j+r}, & \left[
e_{i},e_{\alpha}\right]  =fe_{i}+ge_{i+r}+he_{\alpha}.
\end{array}
\]
Thus, the two vector fields $e_{i}$ and $e_{i+r}$, belonging to the same
eigenvalue $\lambda_{i}$, satisfy the relation:
\begin{equation}
\left[  e_{i},e_{i+r}\right]  =c_{i}e_{i}+c_{i+r}e_{i+r},\label{bII,4,11n}%
\end{equation}
Therefore, $\forall i\in\left\{  1,\cdots,r\right\}  $ the vector fields
$e_{i}$, $e_{i+r}$ are a local basis of a $2-$dimensional involutive
distribution and, by Frobenius' theorem, define a $2-$dimensional submanifold
of ${\cal M}$. In other words, they can be chosen so that, on each
bidimensional manifold, coordinates $\xi^{i},\eta^{i}$ can be found such that
\begin{equation}
e_{i}=\frac{\partial}{\partial\xi^{i}},\quad e_{i+r}=\frac{\partial}%
{\partial\eta^{i}}%
\end{equation}

In conclusion, the relations (\ref{bII,4,10}), which directly follows from the
Nijenhuis condition, ensure the ''partial'' holonomicity of the basis, in
which the tensor field $T$ is diagonal.

On the other hand, since
\begin{equation}
d\lambda_{i}=\vartheta^{j}e_{j}\left(  \lambda_{i}\right)  +\vartheta^{\alpha
}e_{\alpha}\left(  \lambda_{i}\right)  =\vartheta^{j}e_{j}\left(  \lambda
_{i}\right)  ,
\end{equation}
we have, by using Eq. (\ref{bII,4,10}),
\begin{equation}
Td\lambda_{i}=T\vartheta^{j}e_{j}\left(  \lambda_{i}\right)  =\vartheta
^{j}\lambda_{j}e_{j}\left(  \lambda_{i}\right)  =\vartheta^{j}\lambda_{i}%
e_{j}\left(  \lambda_{i}\right)  =\lambda_{i}d\lambda_{i}.
\end{equation}

Moreover,\thinspace\
\[
d\mu^{\rho}\equiv d\mu_{\rho}=\sum_{k=1}^{2r}\vartheta^{i}e_{i}\left(
\mu_{\rho}\right)  +\sum_{\alpha=1}^{2n}\vartheta^{\alpha}e_{\alpha}\left(
\mu_{\rho}\right)  =\sum_{\alpha=1}^{2n}\vartheta^{\alpha}e_{\alpha}\left(
\mu_{\rho}\right)  ,
\]
By means of the above relations, it is now possible to choose a holonomic
basis in such a way as $T$ has the following expression
\begin{equation}
T=\sum_{j=1}^{r}\lambda_{j}\left(  e_{j}\otimes\vartheta^{j}+e_{r+j}\otimes
d\lambda^{j}\right)  \,\,+C_{\rho}^{\sigma}e_{\sigma}\otimes d\mu^{\rho}\,,
\end{equation}
with
\[
C_{\rho}^{\sigma}=\sum_{\alpha=2r+1}^{2n}\mu_{\alpha}e_{\alpha}\left(
\mu^{\sigma}\right)  \left[  e_{\alpha}\left(  \mu^{\rho}\right)  \right]
^{-1}\qquad and\qquad\vartheta^{i}=0.
\]

In addition, in a neighborhood of each bidimensional submanifold we can choose
coordinates $\left(  \lambda/\chi/\mu\right)  $ such that the tensor $T$ can
also be written in the form\footnote{The symbols used for the coordinates have
been choosen just to correspond to the geometric structures previously
described.}:
\[
T=\sum_{j=1}^{r}\lambda_{j}\left(  \frac{\partial}{\partial\lambda_{i}}\otimes
d\lambda_{i}+\frac{\partial}{\partial\chi^{i}}\otimes d\chi^{i}\right)
\,\,+C_{\rho}^{\sigma}\frac{\partial}{\partial\mu_{\rho}}\otimes d\mu_{\sigma
}.
\]
In the chosen basis, the vector field $\Delta$ can be written as
\begin{equation}
\Delta=\Lambda_{i}\frac{\partial}{\partial\lambda_{i}}+\Phi^{i}\frac{\partial
}{\partial\chi^{i}}+E^{\alpha}e_{\alpha},
\end{equation}
so that the condition $L_{\Delta}T=0$ implies that $\Lambda_{i}=E^{\alpha}=0$.
It follows that
\begin{equation}
\Delta=\Phi^{i}\left(  \lambda_{i},\chi^{i}\right)  \frac{\partial}%
{\partial\chi^{i}}.
\end{equation}

Symplectic structures can be found with respect to which the above vector
field is Hamiltonian. Indeed, the closed $2$-form%

\[
\omega=\sum_{k=1}^{r}G_{k}\left(  \lambda_{k},\chi^{k}\right)  d\lambda
_{k}\wedge d\chi^{k}+\sum_{\alpha,\beta=2r+1}^{2n}f_{\alpha\beta}\left(
\mu_{\alpha},\mu_{\beta}\right)  d\mu_{\alpha}\wedge d\mu_{\beta},
\]
will be invariant if
\[
\frac{\partial}{\partial\chi^{i}}\left(  G_{i}\Phi^{i}\right)  =0.
\]
The non degeneracy condition for \ $\omega$ is obviously expressed by
\[
\det\left\|  f_{\alpha\beta}\right\|  \prod_{k=1}^{r}G_{k}\neq0.
\]
This is equivalent to require that if $\Phi^{i}\left(  \lambda_{i},\chi
^{i}\right)  $ vanishes at some point then it also vanishes on the whole
integral curve of \ $\frac{\partial}{\partial\chi^{i}}$ \ through that point.

If the vector field $\Delta$ has no singular points\footnote{If $\Phi_{k}$ is
identically zero for some index $k$, we can define
\[
\omega=\sum_{i}g_{i}\left(  \lambda_{i}\right)  d\lambda_{i}\wedge d\chi
_{i}+\sum_{j}\frac{g_{j}\left(  \lambda_{j}\right)  }{\Phi^{j}\left(
\lambda_{j},\chi_{j}\right)  }d\lambda_{j}\wedge d\chi_{j}+\sum_{\alpha
,\beta=2r+1}^{2n}f_{\alpha\beta}\left(  \mu_{\alpha},\mu_{\beta}\right)
d\mu_{\alpha}\wedge d\mu_{\beta},
\]
\par
where the sum on the index $i$ runs over those eigenspaces for which $\Phi
^{j}=0$.
\par
When $\Delta$ has zeroes but does not vanish identically, we have to exclude
this closed subset from our considerations. These sets will be invariant under
the flow so that our analysis can be carried over in the same fashion as we
have done on the complement.}, a particularly simple class of symplectic
structures with respect to which it is Hamiltonian is given by%

\[
\omega=\sum_{k=1}^{r}\frac{g_{k}\left(  \lambda_{k}\right)  }{\Phi^{k}\left(
\lambda_{k},\chi^{k}\right)  }d\lambda_{k}\wedge d\chi_{k}+\sum_{\alpha
,\beta=2r+1}^{2n}f_{\alpha\beta}\left(  \mu_{\alpha},\mu_{\beta}\right)
d\mu_{\alpha}\wedge d\mu_{\beta},
\]
where $g_{k}$ and $f_{\alpha\beta}$ are arbitrary functions such that
\[
\det\left\|  f_{\alpha\beta}\right\|  \prod_{k=1}^{r}\frac{g_{k}}{\Phi^{k}%
}\neq0.
\]

If the submanifold $\mu=$ {\it  const} is compact and connected, we can
introduce, as usual, action-angle coordinates $\left(  J,\varphi\right)  $ so
that the vector field $\Delta$ and the symplectic structure $\omega$, in the
coordinates $\left(  J,\varphi,\mu\right)  $ take the following form:
\[
\Delta=\Delta^{i}\left(  J_{i}\right)  \frac{\partial}{\partial\varphi^{i}}%
\]%
\[
\omega=\sum_{k=1}^{r}f_{k}\left(  J_{k}\right)  dJ_{k}\wedge d\varphi^{k}%
+\sum_{\alpha,\beta=2r+1}^{2n}f_{\alpha\beta}\left(  \mu_{\alpha},\mu_{\beta
}\right)  d\mu_{\alpha}\wedge d\mu_{\beta}.
\]

In this case, the family of symplectic structures with respect to which
$\Delta$ is Hamiltonian is exhaustively described in \cite{Bo96,FR98}. The
tensor field $T$ can be used to generate compatible invariant symplectic
structures according to
\[
\omega_{T}\left(  X,Y\right)  =\omega_{1}\left(  TX,Y\right)  +\omega
_{1}\left(  X,TY\right)  +\omega_{2}\left(  X,Y\right)
\]
with
\[
\omega_{1}=\sum_{k=1}^{r}f_{k}\left(  J_{k}\right)  dJ_{k}\wedge d\varphi
^{k};\,\,\,\,\,\,\,\,\,\omega_{2}=\frac{1}{2}\sum_{\alpha,\beta=2r+1}%
^{2n}f_{\alpha\beta}\left(  \mu_{\alpha},\mu_{\beta}\right)  d\mu_{\alpha
}\wedge d\mu_{\beta}.
\]

\subsubsection{From noncommutative integrability to invariant tensor fields.}

Let us suppose to have a non commutative integrable system according to the
theorem MF1. By the integrability analysis, we have the symplectic structure
$\omega=d\lambda_{i}\wedge d\chi^{i}+dp_{\alpha}\wedge dq^{\alpha}$ and the
equations of the motion
\[
\dot{\lambda}_{i}=0,\;\dot{\chi}^{i}=\nu_{i},\;\dot{p}_{\alpha}=0,\,\dot
{q}^{\alpha}=0,\,\,\,\,1\leq\,\,i\leq r,\quad\,\,r+1\leq\alpha\leq n,
\]
or, calling $\mu$ the collection of the $p$'s and $q$'s, more simply
\[
\dot{\lambda}_{i}=0,\qquad \dot{\chi}_{i}=\nu_{i}%
\qquad \dot{\mu}_{\alpha}=0.
\]
It is easily verified that the following tensor field
\[
T=\sum_{j=1}^{r}\lambda_{j}\left(  \frac{\partial}{\partial\lambda_{i}}\otimes
d\lambda_{i}+\frac{\partial}{\partial\chi^{i}}\otimes d\chi^{i}\right)
\,\,+C_{\rho}^{\sigma}\left(  \mu\right)  \frac{\partial}{\partial\mu_{\rho}%
}\otimes d\mu_{\sigma}.
\]
is invariant and, for all diagonalizable matrix $C_{\rho}^{\sigma}\left(
\mu\right)  =\delta_{\rho}^{\sigma}\mu_{\sigma}$, has a vanishing torsion,
provided that the Hamiltonian function is separable in the form:
\[
H=K_{1}\left(  \lambda\right)  +K_{2}\left(  \mu\right)  ,
\]
with
\[
K_{1}\left(  \lambda\right)  =\sum_{i=1}^{r}H_{i}\left(  \lambda_{i}\right)
\]

If $K_{1}$ is not separable but
\[
\det\left(  \frac{\partial^{2}K_{1}}{\partial\lambda_{j}\partial\lambda_{i}%
}\right)  \neq0,
\]
the construction of the invariant tensor field follows strictly the lines of
section \ref{inv}.

This shows that also in the noncommutative case an invariant torsionless
tensor field can be always found. Of course, such a tensor field always
generates, by repeated application, Abelian algebras of symmetries. Regardless
of the vanishing of the torsion on the whole space, the noncommutative
features are linked to the non degenerate eigenvalues and, then, are still
described by the term $C_{\rho}^{\sigma}\left(  \mu\right)  \frac{\partial
}{\partial\mu_{\rho}}\otimes d\mu_{\sigma}$.

\vskip.5cm

\noindent{\bf Example}
{\it The Kepler dynamics}

\begin{itemize}
\item {\bf A Recursion operator in the commutative case.}

The vector field for the Kepler problem, in spherical-polar coordinates, for
$\Re^{3}-\{0\}$, is globally Hamiltonian with respect to the symplectic form:
\begin{equation}
\omega=\sum_{i}dp_{i}\wedge dq^{i}
\qquad
i=r,\vartheta,\varphi\label{s4}%
\end{equation}
with Hamiltonian $H$ given by:
\begin{equation}
H=\frac{1}{2m}(p_{r}^{2}+\frac{p_{\vartheta}^{2}}{r^{2}}+\frac{p_{\varphi}%
^{2}}{r^{2}\sin^{2}\vartheta})+V(r),~~~~~~V(r)=-\frac{k}{r}%
\end{equation}
In action-angle coordinates $(J,\varphi)$, the Kepler Hamiltonian $H$, the
symplectic form $\omega$ and the vector field $\Delta$ become:
\begin{eqnarray}
H  & =-\frac{mk^{2}}{\left(  J_{r}+J_{\vartheta}+J_{\varphi}\right)  ^{2}%
}\nonumber\\
\omega & =\sum_{h}dJ_{h}\wedge d\varphi^{h}\nonumber\\
\Delta & =\frac{2mk^{2}}{\left(  J_{r}+J_{\vartheta}+J_{\varphi}\right)  ^{3}%
}\left(  \frac{\partial}{\partial\varphi^{1}}+\frac{\partial}{\partial
\varphi^{2}}+\frac{\partial}{\partial\varphi^{3}}\right)
\end{eqnarray}
It has been shown \cite{MV92} that the vector field $\Delta$ is globally
Hamiltonian also with respect to the symplectic form $\omega_{1}$ :
\begin{equation}
\omega_{1}=\sum_{hk}\left.  S^{h}\right.  _{k}dJ_{h}\wedge d\varphi
^{k}\nonumber
\end{equation}
where the matrix $S$ is defined by:
\[
S=\frac{1}{2}\left\|
\begin{array}
[c]{ccc}%
J_{1} & J_{2} & J_{3}\\
J_{2}-J_{3} & J_{1}+J_{3} & J_{3}\\
J_{3}-J_{2} & J_{2} & J_{1}+J_{2}%
\end{array}
\right\|  .
\]
We have:
\begin{equation}
\Delta=\{H_{1},~~\}_{1},
\end{equation}
with Hamiltonian $H_{1}$ given by:
\begin{equation}
H_{1}=-\frac{2mk^{2}}{J_{r}+J_{\vartheta}+J_{\varphi}}%
\end{equation}
and the new Poisson brackets
\begin{equation}
\{f,g\}_{1}=\sum_{hk}\left.  (S^{-1})_{h}\right.  ^{k}\left(  \frac{\partial
f}{\partial J_{h}}\frac{\partial g}{\partial\varphi^{k}}-\frac{\partial
f}{\partial\varphi^{k}}\frac{\partial g}{\partial J_{h}}\right)
\end{equation}
In the original coordinates $(p,q)$ the symplectic form $\omega_{1}$ is simply
written as:
\begin{equation}
\omega_{1}=\sum_{i}dK_{i}\wedge d\alpha^{i}\label{s12}%
\end{equation}
where the functions $K_{i}(p,q)$ and $\alpha^{i}(p,q)$, defined by:
\[
\left\{
\begin{array}
[c]{l}%
K_{1}={\frac{1}{4}}[J_{1}^{2}+(J_{2}-J_{3})^{2}](p,q)\\
K_{1}={\frac{1}{4}}[J_{1}^{2}+(J_{2}-J_{3})^{2}](p,q)\\
K_{3}={\frac{1}{2}}J_{3}[J_{1}+J_{2}](p,q)\\
\alpha^{i}=\varphi^{i}(p,q)
\end{array}
\right.
\]
are considered as functions of $p,q$ by means of the map $J_{i}=J_{i}%
(p,q),\varphi^{i}=\varphi^{i}(p,q)$. As a consequence a mixed invariant tensor
field $T$, defined, for non degenerate $\omega$, by: $\omega(\widehat
{T}X,Y)=\omega_{1}(X,Y)$ can be constructed.

The tensor field
\begin{equation}
T=\sum_{hk}(\left.  S^{h}\right.  _{k}dJ_{h}\otimes\frac{\partial}{\partial
J_{k}}+\left.  \left(  S^{+}\right)  _{h}\right.  ^{k}d\varphi^{h}\otimes
\frac{\partial}{\partial\varphi^{k}})
\end{equation}
has double degenerate eigenvalues and vanishing Nijenhuis torsion, the last
property being equivalent to the compatibility of the symplectic structures
$\omega$ and $\omega_{1}$.

\item {\bf A Recursion operator in the non commutative case.}

The Kepler dynamics has five first integrals given by the components of the
angular momentum and the components of the orthogonal Laplace-Runge-Lenz
vector.

In action-angle coordinates $\left(  J/\varphi\right)  $ such first integrals
are given by
\[
J_{1},J_{2},J_{3},\varphi_{1}-\varphi_{2},\varphi_{2}-\varphi_{3}.
\]
By using the Delauney action-angle coordinates
\begin{eqnarray*}
I_{1}& =J_{1}+J_{2}+J_{3}\equiv\lambda_{1}\\
I_{2}& =J_{2}+J_{3}\equiv\mu_{3}\\
I_{3}& =J_{3}\equiv\mu_{4}\\
\alpha_{1}& =\varphi_{1}\equiv\chi_{1}\\
\alpha_{2}& =\varphi_{2}-\varphi_{1}\equiv\mu_{5}\\
\alpha_{3}& =\varphi_{3}-\varphi_{2}\equiv\mu_{6},
\end{eqnarray*}
we can construct the invariant torsionless tensor field
\[
T=\lambda_{1}\left(  \frac{\partial}{\partial\lambda_{1}}\otimes d\lambda
_{1}+\frac{\partial}{\partial\chi_{1}}\otimes d\chi_{1}\right)  \,\,+\sum
_{\alpha=3}^{6}\mu_{\alpha}\frac{\partial}{\partial\mu_{\alpha}}\otimes
d\mu_{\alpha}.
\]
\end{itemize}

\section{Conclusion}

It has been shown that also in the non commutative case a criterion of
integrability can be formulated in terms of invariant ''semitorsionless''
$\left(  1,1\right)  $ tensor field in close analogy with the commutative
case. Moreover, it has also been shown that in such cases a new invariant
$\left(  1,1\right)  $ tensor field can be constructed with a vanishing
Nijenhuis torsion. By using either of them, sequences of compatible symplectic
structures can be constructed.%

\end{document}